\documentclass[12pt,preprint]{aastex}

\begin{document}


\title{Earliest detection of the optical afterglow of GRB 030329 
and its variability}


\author{R. Sato\altaffilmark{1}, N. Kawai\altaffilmark{1}, M. Suzuki\altaffilmark{1}, Y. Yatsu\altaffilmark{1}, J. Kataoka\altaffilmark{1}, R. Takagi\altaffilmark{1}, K. Yanagisawa\altaffilmark{2}, \\ and H. Yamaoka\altaffilmark{3}}


\altaffiltext{1}{Tokyo Institute of Technology, Meguro-ku, Tokyo, Japan}
\altaffiltext{2}{Okayama Astrophysical Observatory, Asaguchi-gun, Okayama, Japan}
\altaffiltext{3}{Kyusyu University, Ropponmatsu, Fukuoka, Japan}


\begin{abstract}

We report the earliest detection of an extremely bright optical 
afterglow of the gamma-ray burst (GRB) 030329 
using a 30cm-telescope at Tokyo Institute of Technology (Tokyo, JAPAN). 
Our observation started 67 minutes after the burst, and continued for 
succeeding two nights until the afterglow faded below 
the sensitivity limit of the telescope (approximately 18 mag). 
Combining our data with those reported in GCN Circulars, we find that 
the early afterglow light curve of the first half day is described 
by a broken power-law ($\propto$ $t^{-\alpha}$)
 function with indices 
$\alpha_{1} = 0.88 \pm 0.01$ (0.047$<t<$ $t_{\rm \rm b1}$ days), 
$\alpha_{2} = 1.18 \pm 0.01$ ($t_{\rm \rm b1}$ $<t<$ $t_{\rm \rm b2}$ 
days), and 
$\alpha_{3} = 1.81 \pm 0.04$ ($t_{\rm \rm b2}$ $<t<$ 1.2 days), 
where $t_{\rm \rm b1}$ $\sim 0.26$ days and 
$t_{\rm \rm b2}$ $\sim 0.54$ days, respectively. 
The change of the power-law index at the first break at $t\sim0.26$ days 
 is consistent with that expected from a ``cooling-break'' when the cooling 
frequency crossed the optical band. If the interpretation is correct, 
the decay index before the cooling-break implies a uniform ISM environment.

\end{abstract}




\section{Introduction}

The overall behavior of GRB afterglows has been successfully
explained by the standard fireball model (e.g. Piran et al. 1999).  
It is expected that the deviation from a simple power-law 
in the afterglow light curve will provide wealth of information 
on the environment and physical parameters of GRBs.  
For example, observations of early GRB afterglows confirmed that 
``breaks'' exist in the light curves of a number of GRBs. 
Such breaks may be understood in the framework of the standard fireball
model either as 
a ``jet-break'' where the bulk Lorentz factor of the relativistic jet 
decrease to the inverse of jet opening angle 
(Sari, Piran $\&$ Halpern 1999, Rhoads 1999), or 
the ``cooling-break'' where the high energy electrons start 
to lose most of their energy rapidly by synchrotron emission 
which emit the observed optical photons (Sari, Piran $\&$ Nakar 1998). 
While there are number of convincing cases for jet-breaks,
identification of a cooling-break in the optical light curve has been 
difficult since it requires detection of a subtle change 
in the power-law index ($\Delta \alpha \sim 0.25$).
In order to study afterglow light curves in detail, continuous coverage
and a high signal-to-noise ratio are required.

The situation has been dramatically improved since the advent of 
High Energy Transient Explorer-2 (HETE-2). 
HETE-2 can determine the positions of GRBs onboard, 
and notify ground-based observers of the GRB coordinates  
within 1$-$10 minutes after the burst (Ricker et al. 2002).
For example, the locations of GRB 021004 (e.g., Fox 2002) 
and GRB 021211 (e.g., Fox $\&$ Price 2002) were disseminated 
within less than a minute after the bursts, 
which prompted detailed studies of GRB afterglows in the very initial
phases while they are bright.

GRB 030329 was detected by the HETE-2 satellite on 29 March 2003 at
11:37:14.7 UT. 
Position was determined by the ground analysis, 
and the location was reported to GCN at 73 minutes after the burst 
(Vanderspek et al. 2003). A very bright($R \sim 13$ mag) optical 
transient (OT) was reported at $\alpha = 10^{h} 44^{m} 50^{s} .0$, 
$\delta = + 21^{\circ} 31^{\prime} 17.8'' (J2000.0)$ 
(Peterson $\&$ Price 2003, Torii 2003) inside the SXC error circle. 
This is the brightest GRB ever detected by HETE-2 with the 30$-$400 keV 
fluence of $1.2 \times 10^{-4}$ erg cm$^{-2}$, and precise and continuous 
follow-up observations were carried out by dozens of telescopes 
distributed around the world. 
Optical spectroscopic observations have determined its redshift as 
$z$ = 0.1685 (Greiner et al. 2003), which is one of the closest ever known
and is possibly related to the exceptional brightness of this afterglow. 
Moreover, spectra taken after several days reveal the evolution of broad 
peaks in the spectra characteristic of a supernova.
The spectral similarity to SN 1998bw (e.g., Galama et al. 1998, 
Iwamoto et al. 1998) and other energetic supernovae such as 1997ef 
provides strong evidence that GRB 030329 is associated with 
core-collapse supernovae 
(Dado, Dar $\&$ Rujula 2003, Hjorth et al. 2003, Stanek et al. 2003, 
Kawabata et al. 2003).
In order to investigate the kinetic energy of GRB and the immediate
vicinity of its progenitor, the early light curve is important.  
In this letter, we report the earliest detection of the optical afterglow 
of GRB 030329 starting 67 minutes after the burst.

\section{Observation and photometry}

Our observation was performed at Tokyo Institute of Technology using 
a 30cm-telescope (Meade LX-200) and an unfiltered CCD camera (Apogee AP6E) 
equipped with a front-illuminated 1024$\times$1024 CCD chip (Kodak KAF-1001E). 
The dark current images were subtracted from the obtained CCD images and 
then flat fielding was applied for all images. 
We used IRAF/noao/digiphot/apphot/phot packages to analyze the data.

We started observing the preliminary SXC position at 12:44:13 
UT on 29 March 2003, 67 minutes (0.047 days) after the burst
\footnote{As a HETE-2 Ops graduate student R. Sato took the initiative 
and ``ran up to the roof to start observing'' while the location data 
were still preliminary.}. 
The magnitude of GRB afterglow at the very beginning was $R_{c} \sim 12.4$ 
mag. 
This is the earliest detection of the afterglow of GRB 030329 
ever reported in literature. 

We continued observations for the rest of the night covering 
$t \sim 0.05-0.30$ days, and performed observations on the following 
two nights covering the period of $t \sim 0.93-1.21$ days 
and $t \sim 2.03-2.08$ days, respectively, 
where $t$ refers to the time since the burst onset. 
The exposure time of each CCD frame was 10 sec ($0.047 < t <1.21$ days; 
when the after glow was relatively bright) or 30 sec (2.03 days$< t$). 
The magnitude of the comparison stars were calibrated using three stars 
in the same field of view, which has been calibrated in detail by Henden. Then we determined the magnitude of the OT in each frame relative to 
the weighted average of 15 bright comparison stars.

Since the peak sensitivity of our camera is very close to the $R_{c}$ band, 
we have calibrated magnitude of the OT by converting our magnitude system
($R_{inst}$) into the $R_{c}$ system assuming the color correction with 
the color $(V-I)$ = 0.74 mag (Zeh et al. 2003) ($a(r) = (R_{inst}) - 
0.1514 \times (V-I)$, 
where a(r) is the zero-point between the instrumental system and 
the standard magnitude). 
The statistical error is $\sim 0.017-0.02$ mag.
Although the zero-point errors were found to be $\sim$ 0.03 mag 
for the reference stars by Henden (2003), 
we found our data were consistently brighter than the $R$-band 
filtered observations (e.g., Burenin et al. 2003). 
We therefore introduced additional zero-point correction (+0.11 mag) 
to match the light curves in the overlapped interval.
The resulting light curve of the GRB 030329 afterglow in the $R_{c}$ band is 
shown by combining data from other observations (see Fig 1). 

The light curve in the first day cannot be fitted with a single 
power-law function. 
We therefore tried to fit the light curve by two different forms of 
broken power-law functions. One is given by Beuermann et al.:
$F(t) \propto	
{ [ (t / t_{b})^{\alpha_{1}^{\prime} n^{\prime}} + 
(t / t_{b})^{\alpha_{2}^{\prime} n^{\prime}} ]^{-1 / n^{\prime}} }$,
where $t_{b}$ is the break time, and $n^{\prime}$ provides a measure of 
the relative width and the smoothness of the break.
The other is a ``double-broken power-law'' function with two breaks 
with the following form:

\begin{equation}
F(t) \propto \left\{ 
\begin{array}{@{\,}ll}
t^{-\alpha_{1}} & \mbox{ ( $t < t_{\rm b1}$ ) } \\
$[$ (t / t_{b2})^{\alpha_{2} n} + (t / t_{b2})^{\alpha_{3} n} $]$^{-1 / n} 
& \mbox{ ( $t > t_{\rm b1}$ ) }
\end{array}
\right .\
\end{equation}

where $t_{\rm b1}$ and $t_{\rm b2}$ are 
the break times and $n$ provides a measure of the relative width 
and the smoothness of the break. 
Here we excluded the ``bump'' at $t \sim 0.08 - 0.09$ days, 
which is discussed in section 3.3. 

We found that the former is not acceptable with a reduced $\chi^{2}$ 
of 1.72 (285 d.o.f) whereas the latter improves the fit significantly 
(the reduced $\chi^{2}$ 1.06 with 283 d.o.f) (see Fig 2). 

%The light curve of GRB 030329 cannot be fitted with a single power-law 
%function. 
%So we tried fitting the light curve by two types of a broken power-law. 
%One is assumed there is only a jet-break at $t_{\rm b2}$, 
%another is assumed there are two breaks at $t_{\rm b1}$ and $t_{\rm b2}$, 
%where $t_{\rm \rm b1} \sim 0.26$ days and $t_{\rm \rm b2} \sim 0.54$ days, 
%respectively.  
%When two results are compared, it turned out that the model considered 
%as there being two breaks is fitted best (see caption in Fig 2). 
      
As a result, it is well described by a broken power-law of the form; 
$\alpha_{1} = 0.88 \pm 0.01$ 
($0.047 < t < t_{\rm b1}$ days), 
$\alpha_{2} = 1.18 \pm 0.01$ 
($t_{\rm b1} < t < t_{\rm b2}$ days),
$\alpha_{3} = 1.81 \pm 0.04$ 
($t_{\rm b2} < t < 1.2$ days), where $t_{\rm b1} \sim 0.26$ days 
and $t_{\rm b2} \sim 0.54$ days, respectively, and $n = 18.8 \pm 5.1$. 
Here, $\alpha_{1}$ is determined by essentially the full Tokyo Tech data. 
The earliest phase of the light curve is well fitted by the single power-law
 with its index $\alpha_{1}$.
%although there is a sign of an excess (``bump'') 
%in the light curve at $t \sim 0.08-0.09$ days.
$\alpha_{2}$ and $\alpha_{3}$ are determined by measurements 
reported by Burenin et al. and GCN (see caption in Fig 1). 

\section{Discussion}

\subsection{Light curve at $0.05 < t < 0.26$ days}

We have presented a light curve of the early phase of the optical afterglow 
of GRB 030329 starting 67 minutes after the burst. 
This is the earliest detection of GRB 030329 ever reported 
(Peterson $\&$ Price 2003, Torii et al. 2003, Uemura et al. 2003). 

Burenin et al. (2003) reported follow-up observations of GRB 030329 
as early as 6 hours after the burst, using BVRI filters. 
In each of the filters, they observed a gradual flux decay 
which can be accurately described as a power-law 
$F_{\nu}$ $\propto$ $t^{-1.19}$. They also found a characteristic 
break in the light curve $t_{\rm brk}$ $\sim$ 0.57 day, 
after which the afterglow flux started to decline faster.  
The power-law slopes of the light curves changed from $-$1.19 to $-$1.9 
for $t \ge t_{\rm brk}$. 
Notably, this break is nicely consistent with our second break 
($t_{\rm b2}$) within error bars, where the power-law slopes changes 
from $-1.18 \pm 0.01$ to $-1.81 \pm 0.04$ after the break of 
$t_{\rm b2} \sim 0.54$ days (see above).  
The first break ($t_{\rm b1}$) is not discussed in Burenin et al,
since they started their observations just around this break 
($t_{\rm b1} \sim 0.26$ days).

Price et al. (2003), Burenin et al. (2003), and Tiengo et al. (2003) 
found that the results of their observations are consistent 
with the model where the afterglow emission is generated during the 
deceleration of the ultra-relativistic collimated jet. 
They found that the break in the power-law light curve, 
at $t\sim0.5-0.6$ days, can be interpreted as the jet-break, 
i.e., the break which occurs when the angular structure 
of the ultra-relativistic collimated jet becomes observable 
(Sari, Piran $\&$ Halpern 1999, Rhoads 1999).
This interpretation is supported by the fact that the break occurred 
simultaneously in different colors. 
Furthermore, a change in power-law slope from $-$1.19 to $-$1.9, 
is approximately consistent with that generally observed in jet-break. 
Therefore, our major concern in this paper is to understand the nature 
of the first break, $t_{\rm b1}$, and examine the consistency 
between the above scenario in the frame work of standard GRB fireball theories 
(Piran et al. 1999).
\vspace*{-0.3cm}

\subsection{Break at $t \sim 0.26$ days}

There are two possible break frequencies in the spectra,  
 $\nu_{\rm m}$ and $\nu_{\rm c}$, where 
$\nu_{\rm m}$ is the synchrotron frequency, 
$\nu_{\rm c}$ is the cooling frequency above which electrons lose 
their energy rapidly by synchrotron radiation (Sari, Piran $\&$ Narayan 1998). 
Since $\nu_{\rm m}$  and $\nu_{\rm c}$ are the functions of time, 
a break in the light curve could be observed when $\nu_{\rm c}$ and/or 
$\nu_{\rm m}$ crossed over the observed frequency $\nu_{\rm R}$. 
Therefore, we examined six possible cases to understand the first break 
($t_{\rm b1}$), according to the relation between  $\nu_{\rm m}$, 
$\nu_{\rm c}$ and $\nu_{\rm R}$. 
In the standard GRB scenario, $\nu_{\rm c}$ $\le$ $\nu_{\rm m}$ is 
often called ``fast cooling'' since all electrons cool rapidly, 
whereas $\nu_{\rm c}$ $\ge$ $\nu_{\rm m}$   is referred to ``slow cooling'' 
since only the high energy population of electrons cool efficiently. 
We will also extend our discussion to discriminate between 
``a homogeneous interstellar medium (ISM) model'' (e.g., Sari $\&$ Piran 1999)  and  ``a pre-existing stellar wind model'' (e.g., Chevalier
$\&$ Li 1999) for the GRB environment. 
The relationships between observed spectral index and 
model predictions are compared in Table 1.

We first consider the case where both $\nu_{\rm c}$ and $\nu_{\rm m}$ are 
above the observed optical frequencies  ($\nu_{\rm R} < \nu_{\rm m} < \nu_{\rm c}$ or $\nu_{\rm R} < \nu_{\rm c} < \nu_{\rm m}$: case (3) and (6) 
in Table 1). 
In these two situations, observed flux at $\nu_{\rm R}$ should increase 
with time, which strongly conflicts with the observed declining light curve. 
On the contrary, if the both cooling frequency $\nu_{\rm c}$ 
and the minimum frequency $\nu_{\rm m}$ are below the optical band 
($\nu_{\rm c} < \nu_{\rm m} < \nu_{\rm R}$ or $\nu_{\rm m} < \nu_{\rm c} < 
\nu_{\rm R}$: case (1) and (4) in Table 1), the predicted optical spectral index would 
be $\beta = p/2$, where $p$ is the electron spectral index. 
Since the photon spectral index of this afterglow was $\beta = 0.66$ 
at $t = 0.26$ days (Burenin et al. 2003), 
we expect $p = 1.32$, which is unusually flat for an electron population 
accelerated in a GRB. 
Furthermore, power-law index in the light curve should be 
$\alpha = \frac{2-3p}{4} \sim 0.49$, which is again inconsistent 
with $\alpha_{1} \sim 0.88$ obtained with our data. 

Case (5) $\nu_{\rm c} < \nu_{\rm R} < \nu_{\rm m}$ in Table 1 is also  
ruled out since the predicted power-law index $\alpha = 0.25$ 
(Sari, Piran $\&$ Narayan, 1998) is too flat compared to the observed value of 
$\alpha_{1} \sim 0.88$. 
Therefore, we argue that the possible solution is 
$\nu_{\rm m} < \nu_{\rm R} < \nu_{\rm c}$ for the time region of 
$t \leq t_{\rm b1}$. 
In this case, however, if the burst occurred in pre-existing stellar wind, 
the optical decay slope is predicted to be $\alpha = \frac{3\beta}{2} - 
\frac{\delta}{8-2\delta} \sim 1.49$, with $\delta = 2$ for a wind model 
(Panaitescu, Meszaros $\&$ Rees 1998), which is quite steeper than 
that observed, and hence we can rule out wind-interaction model.
In summary, $\nu_{\rm m} < \nu_{\rm R} < \nu_{\rm c}$ and ISM model 
(case (2) in Table 1) is the only possible solution to reproduce 
both the temporal/spectral index of the optical afterglow of GRB 030329 
at $0.05 < t < 0.26$ days. 

In such a slow cooling case, time variation of afterglow flux 
is given by $F \propto t^{-3(p-1)/4}$ for $\nu < \nu_{\rm c}$ and 
$F \propto t^{-(3p-2)/4}$  for $\nu_{\rm c} < \nu$, 
respectively (Sari, Piran $\&$ Narayan 1998). 
By assuming $\alpha_{1}$ = 0.88, the electron spectral index is estimated 
as $p = 2.17$.
Note that, this electron spectral index agrees well with those of 
electrons accelerated in relativistic shock waves (e.g., Dado, Dar $\&$ Rujula 2003). Furthermore, we expect that power-law slope of the light curve 
would change from 0.88 to 1.13 for $\nu_{\rm c} < \nu$.
Again, this is approximately consistent with the observed spectral index after 
$t_{\rm b1}$, where $\alpha_{2} = 1.18 \pm 0.01$. 
Therefore, we conclude  
that the first break in the optical afterglow light curve at $t_{\rm b1}$ 
is the most probably cooling-break where the cooling frequency crosses 
down the observed optical frequency. 

Under this assumption, 
we can determine important physical parameters for the GRB emission.
For example, we can estimate $\epsilon_{B}$ and  $\epsilon_{e}$, 
where $\epsilon_{B}$ and $\epsilon_{e}$ are the fractions of the shock 
energy given to magnetic field and electrons at the shock (Sari, Piran $\&$ 
Narayan 1998). 
In case of slow cooling, $t_{m}<t_{R}<t_{c}$ would be expected.
Since we started our observation 0.047 days after the burst, we can limit 
the range of $t_{m}$ as $t_{m} < t_{obs} = 0.047$ days. 
For $t_{c} = t_{\rm b1} = 0.26$ days, $t_{m} < 0.047$ days, 
$E = 10^{52}$ erg, $n = 1$ cm$^{-3}$, $\nu = 0.5 \times 10^{15}$ Hz, 
we obtain

\begin{equation}
\epsilon_{B} \sim  0.05 
\biggl( \frac{t_c}{0.26} \biggr)^{-\frac{1}{3}}
\biggl( \frac{E_{52}}{1.0} \biggr)^{-\frac{1}{3}} 
\biggl( \frac{n_1}{1.0} \biggr)^{-\frac{2}{3}}
\biggl( \frac{\nu_{15}}{0.5} \biggr)^{-\frac{2}{3}} 
\end{equation}

\begin{equation}
\epsilon_{e} < 0.20 
\biggl( \frac{\epsilon_B}{0.05} \biggr)^{-\frac{1}{4}}
\biggl( \frac{t_m}{0.047} \biggr)^{\frac{3}{4}}
\biggl( \frac{E_{52}}{1.0} \biggr)^{-\frac{1}{4}} 
\biggl( \frac{\nu_{15}}{0.5} \biggr)^{\frac{1}{2}} 
\end{equation}

We can also 
constrain the peak time of reverse shock $t_{A}$ days after the burst 
%from the occurrence of the GRB 
(Sari $\&$ Piran 1999). 

\begin{eqnarray}
t_A \sim 0.03
\biggl( \frac{\epsilon_B}{0.05} \times \frac{1}{0.1} \biggr)^{-3}
\biggl( \frac{E_{52}}{1.0} \biggr)^{-1}
\biggl( \frac{n_{1}}{1.0} \times \frac{\nu_{c,15}}{0.5} \biggr)^{-2}  
\end{eqnarray}

%\begin{eqnarray}
%t_A \sim 0.03
%\biggl( \frac{\epsilon_B}{0.05} \times \frac{1}{0.1} \biggr)^{-3}
%\biggl( \frac{E_{52}}{1.0} \biggr)^{-1}
%\biggl( \frac{n_{1}}{1.0} \biggr)^{-2}  \nonumber \\
%\times
%\biggl( \frac{\nu_{c,15}}{0.5} \biggr)^{-2}   
%{\rm days}
%\end{eqnarray}

The Lorentz factor depends on time, 
$\gamma(t) \sim (3E/256 \pi n m_{p}c^{5}t^{3})^{1/8}$ (Piran 1999). 
And the magnetic field is calculated using $\epsilon_{B}$ and the Lorentz
factor by $B = \gamma c \sqrt{32 \pi \epsilon_{B} n m_{p}}$. 
Assuming $E = 10^{52}$erg, the Lorentz factor and the magnetic field strength 
at two characteristic breaks time are $\gamma = 9.7, B = 0.86$ gauss 
at $t_{\rm b1} = 0.26$ days and $\gamma = 7.4, B = 0.64$ gauss 
at $t_{\rm b2} = 0.54$ days, respectively. 

The values of $\epsilon_{B}$ and $\epsilon_{e}$ estimated in the preceding 
chapter are in good agreement with the averages of these parameters for GRBs 
calculated by Panaitescu $\&$ Kumar (2001), which are
$\log \epsilon_{B} = -2.4 \pm 1.2$ and $\epsilon_{e} = 0.062 \pm 0.045$.

\subsection{Bump at $t \sim 0.08-0.09$ days}

Finally, we comment on a small ``bump'' of the light curves 
at $t \sim 0.08-0.09$ days ($t_{\rm bump}$) with an amplitude of $\sim0.1$ mag.
Uemura et al. (2003) reported a change of slopes from 0.74 to 0.95 days 
at $t = 0.085$ days. 
However our earliest data at $t < 0.08$ days has a slope steeper than 0.74. 
The light curve at $t > 0.09$ days lies on the extrapolation 
of this earliest segment. 
We consider this feature as a bump rather than a break. 

Short time variabilities, i.e., ``bumps and wiggles'', may be associated with 
the forward/reverse shock structures along the afterglow emitting regions 
(Kobayashi $\&$ Zhang 2003), repeated energy injection from the central engine, or fluctuation in the density of the interstellar medium 
(Nakar, Piran $\&$ Granot 2003).

First, we can rule out a case with the forward/reverse shock structure, 
since it predicts the light curve should not have the same power-law index
before and after the bump.
A case with repeated energy injection is also ruled out 
since after the injection the light curve after the bump should have
the same power-law decay slope, but with a larger normalization.
Therefore, we conclude that the bumps in the light curve is likely due to 
the fluctuation in the external density of the interstellar medium 
(Nakar, Piran $\&$ Granot 2003).

We can estimate the distance from the central engine $R(t)$ (Piran 1999) and 
the density variation (Nakar, Piran $\&$ Granot 2003) at $t_{\rm bump}$.

\begin{equation}
R(t) \sim 2.2  \times 10^{17} 
\biggl[ 3 
\biggl( \frac{E_{52}}{1.0} \biggr)
\biggl( \frac{t (\rm sec)}{7300} \biggr) 
\bigg/ \pi m_{p} c
\biggl( \frac{n_1}{1.0} \biggr)
\biggr]^{\frac{1}{4}}
{\rm cm}
\end{equation} 

\begin{equation}
(n / n_{0}) \sim 1.1 \times [ (F_{\nu}/F_{0}) / 1.1 ] ^{4/(1+p)}
\end{equation} 

%\begin{equation}
%\biggl( \frac{n}{n_{0}} \biggr) \sim 1.1 \times  
%\biggl( 
%\frac{F_{\nu}/F_{0}}{1.1} 
%\biggr)^{4/(1+p)}     
%\end{equation} 

We find that density is enhanced about 10$\%$ at 
the distance of $2.2 \times 10^{17}$ cm.

\section{Summary}

We observed extremely bright optical afterglow of GRB 030329 from 67
minutes after the burst.
Our observational results show that the shocked electrons are in the
slow cooling regime with an electron index of 2.17 in this burst, and
that the burst occurred in a uniform ISM, that is, GRB 030329 can be
understood very well in the predicted ``standard'' model.
We conclude the first break changes the power law index by $\sim 0.3$,
consistent with the cooling-break in the frame work of the standard
external shock model.

%The present result may contradict the general expectation that
%the progenitor of a GRB, being an extremely massive star, has wind-type
%density environment.
%Clearly, more theoretical and observational studies of GRB and 
%its environment are awaited.  


%% Included in this acknowledgments section are examples of the
%% AASTeX hypertext markup commands. Use \url without the optional [HREF]
%% argument when you want to print the url directly in the text. Otherwise,
%% use either \url or \anchor, with the HREF as the first argument and the
%% text to be printed in the second.

\acknowledgments

We are grateful to the other members of the HETE-2 Ops Team for
providing the location of GRB 030329. We are also grateful to 
R. Burenin and his collaborators for kindly providing us 
the numerical values of the filtered data and to K. Torii for giving 
useful advises on photometry. 
This work is supported by the Grants-in-aid for Scientific Research Program 
from the Ministry of Education, Science and Culture of Japan to NK.

\clearpage


\begin{figure}
\plotone{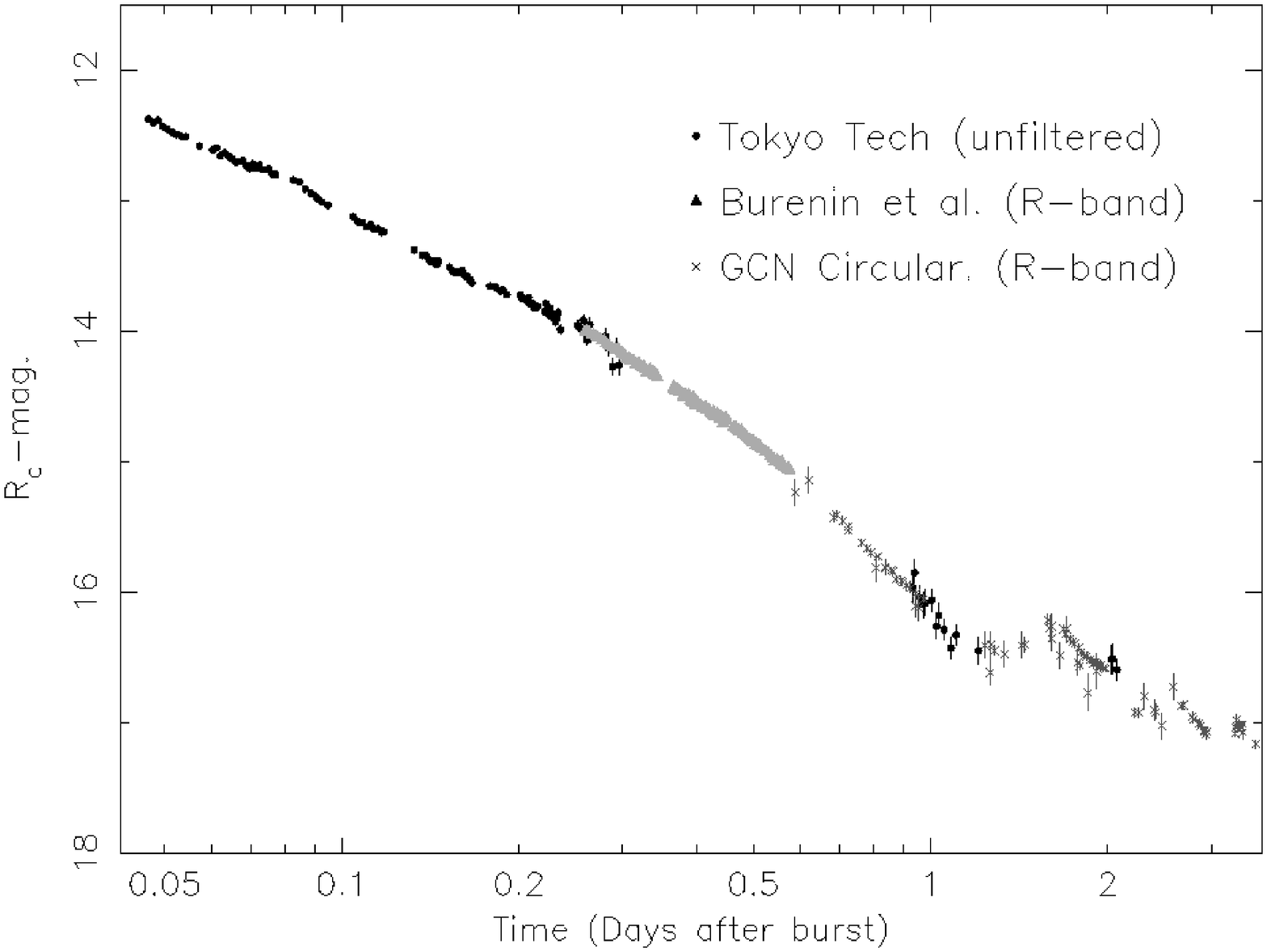}
\caption{Light curve of the optical afterglow of GRB 030329. The filled circles are our observations, the filled triangles come from Burenin et al. (2003), and the rest comes from GCN 2028, 2029, 2034, 2041, 2050, 2056, 2058, 2064, 2067, 2070, 2071, 2074, 2077, and KAIT. These magnitudes were translated using the standard sequence by Henden (2003). \label{f1}}
\end{figure}

\clearpage

\begin{figure}
\plotone{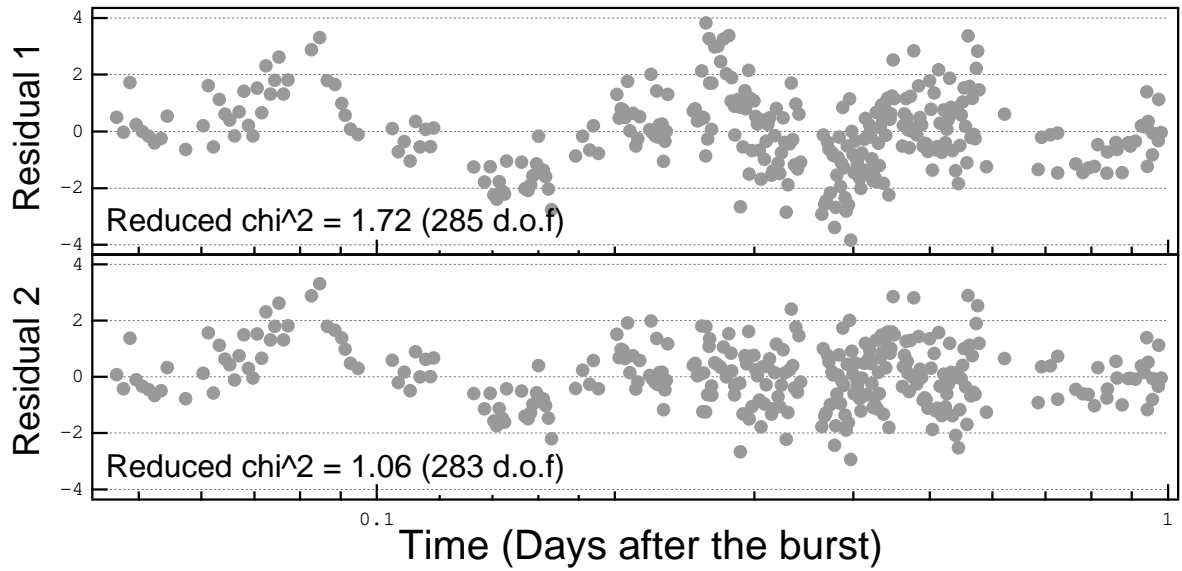}
\caption{Residuals of the light curve of the optical afterglow for the two models. $top$; the residual from a broken power-law which is containing a single, broad break. $bottom$; the residual from a double broken power-law (see text). \label{f2}}
\end{figure}

\clearpage 

%% If you are not including electonic art with your submission, you may
%% mark up your captions using the \figcaption command. See the 
%% No more than seven \figcaption commands are allowed per page, 
%% so if you have more than seven captions, insert a \clearpage 

%% Tables should be submitted one per page, so put a \clearpage before

%% Two options are available to the author for producing tables:  the
%% deluxetable environment provided by the AASTeX package or the LaTeX
%% table environment.  Use of deluxetable is preferred.
%%

%% Three table samples follow, two marked up in the deluxetable environment,
%% one marked up as a LaTeX table.

%% In this first example, note that the \tabletypesize{}
%% command has been used to reduce the font size of the table.
%% Note also that the \label command needs to be placed 

\clearpage

\begin{deluxetable}{clccl}
\tabletypesize{\scriptsize}
\tablecaption{\footnotesize Predicted decay slopes for various theoretical models}
\tablewidth{0pt}
\tablehead{
\colhead{Model} &
\colhead{Environment} &
\colhead{$\alpha$} &
\colhead{Comment} 
}
\startdata
(1)$\nu_{\rm m}<\nu_{\rm c}<\nu_{\rm R}$ &
 {ISM}  & 0.49 & $\alpha$ and $\beta$ are inconsistent \\
 &                     
 {Wind} & 0.49 & $\alpha$ and $\beta$ are inconsistent \\
(2)$\nu_{\rm m}<\nu_{\rm R}<\nu_{\rm c}$ &
 {ISM}  & 0.99 &  O.K.                                 \\
 &                      
 {Wind} & 1.49 & $\alpha$ does not fit data            \\
(3)$\nu_{\rm R}<\nu_{\rm m}<\nu_{\rm c}$ & 
 {ISM}  &  -   & $\alpha<0$                            \\
 &                      
 {Wind} &  -   & $\alpha<0$                            \\
(4)$\nu_{\rm c}<\nu_{\rm m}<\nu_{\rm R}$ & 
 {ISM}  & 0.49   & $\alpha$ and $\beta$  are inconsistent \\      & 
 {Wind} & 0.49   & $\alpha$ and $\beta$  are inconsistent \\
(5)$\nu_{\rm c}<\nu_{\rm R}<\nu_{\rm m}$ & 
 {ISM}  & 0.25 & $\alpha$ does not fit data            \\
 &
 {Wind} & 0.25 & $\alpha$ does not fit data            \\
(6)$\nu_{\rm R}<\nu_{\rm c}<\nu_{\rm m}$ & 
 {ISM}  &  -   & $\alpha<0$                            \\ 
 &
 {Wind} &  -   & $\alpha<0$                            \\
\enddata
\end{deluxetable}

%% Text for table notes should follow after the \enddata but before
% the \end{deluxetable}. Make sure there is at least one \tablenotemark
%% in the table for each \tablenotetext.

\end{document}